# Single-walled and multiwalled carbon nanotubes viewed as elastic tubes with the effective Young's moduli dependent on layer number

Zhan-chun Tu*
*Institute of Theoretical Physics, The Chinese Academy of Sciences, P.O. Box 2735 Beijing 100080, China*

Zhong-can Ou-Yang
*Institute of Theoretical Physics, The Chinese Academy of Sciences, P.O. Box 2735 Beijing 100080, China
and Center for Advanced Study, Tsinghua University, Beijing 100084, China*

The complete energy expression of a deformed single-walled carbon nanotube (SWNT) is derived in the continuum limit from the local density approximation model proposed by Lenosky et al. [Nature (London) **355**, 333 (1992)] and is shown to be consistent with the classic shell theory by which Young's modulus, the Poisson ratio, and the effective wall thickness of SWNT's are obtained as $Y=4.70$ TPa, $\nu=0.34$, and $h=0.75$ Å, respectively. The elasticity of a multiwalled carbon nanotube (MWNT) is investigated as the combination of the above SWNT's of layer distance $d=3.4$ Å and the effective Young's modulus of the MWNT is found to be an apparent function of the number of layers, $N$, varying from 4.70 to 1.04 TPa for $N=1$ to $\infty$.

Since carbon nanotubes (CNT's) were discovered in 1991,[1] their mechanical properties have been the subject of a number of theoretical[2–8] as well as experimental[9–14] studies. CNT's are predicted to be of high stiffness,[2–11] high axial strength,[3,4,10,12] strong flexibility,[13] and low friction.[14] As is well known, a single-walled carbon nanotube (SWNT) can be thought of as only one graphitic layer with hexagonal lattices that has been wrapped up into a seamless cylinder.[15] Thus this raises a question as to how to determine "the wall thickness" for a SWNT in an explicit way. The puzzle has led a wide predicted range (0.5–5.5 TPa) of Young's moduli $Y$ for the SWNT.[2–8] Using molecular dynamics simulations and fitting them for the elastic shell theory, Yakobson et al.[3] calculated the Poisson ratio $\nu=0.19$, effective wall thickness $h=0.66$ Å, and $Y=5.5$ TPa, respectively. Selecting $\nu=0.24$ and using the tight-binding method and elastic theory, Zhou et al.[6] predicted that the effective wall thickness and Young's modulus of SWNTs are $h=0.74$ Å and $Y=5.1$ TPa. On the other hand, the using empirical force-constant method and adopting an *ad hoc* convention $h=3.4$ Å which is the interlayer distance of graphite, Lu[8] obtained $\nu=0.28$ and $Y=1$ TPa. Recently, Krishnan et al.[11] adopted the same convention and measured the average value of $\langle Y \rangle = 1.25 \pm 0.4$ TPa (i.e., $\langle Y \rangle = 5.7 \pm 1.8$ TPa if taking $h=0.74$ Å). In fact, the *ad hoc* convention is possible only in the treatment of a simple stretch deformation of the SWNT. For a "curved SWNT" the values of $h$ and $Y$ can be separately obtained.[6] Thus a natural question for the SWNT arises as follows: what is the unique definition of the wall thickness of the SWNT and can we derive it as well as $Y$ and $\nu$?

To the Young's modulus of multiwalled carbon nanotubes (MWNT's), the theoretical work is relatively lacking. Lu[8] and other authors[7] have estimated their Young's modulus to be about 1 TPa, which is close to the modulus of graphite but much smaller than the experimental values $1.8 \pm 0.9$ TPa (Ref. 9) and $1.3 \pm 0.6$ TPa (Ref. 10). The experimental errors are too large. Maybe there are some unknown and underlying reasons.

Moreover, the elastic properties of SWNT's and MWNT's belong to the domain of nanomechanics, which is between micromechanics (i.e., quantum mechanics) and macromechanics (e.g., classic elastic theory). The treatment of them is divided into two steps: first, calculate the variation of microscopic electronic energies caused by the structure deformations,[16] and second, derive the corresponding deformation energy with classic elastic theory and obtain the elastic constants such as the Poisson ratio, Young's modulus, and so on from the relation between the two energies. However, is it always possible to do the procedure? In other words, is it equivalent for both energies in form?

Here we report a new resolution of the above questions. The complete energy expression of a deformed SWNT is derived in the continuum limit from the local density approximation model proposed by Lenosky et al.[17] and can be well expressed as the form of classical elastic theory with Young's modulus $Y=4.70$ TPa, Poisson ratio $\nu=0.34$, and effective wall thickness $h=0.75$ Å. The elasticity of a MWNT is investigated as the combination of the above SWNTs of layer distance $d=3.4$ Å and the effective Young's modulus of the MWNT is found to be an apparent function of the number of walls, $N$, varying from 4.70 to 1.04 TPa for $N=1$ to $\infty$.

We start from the concise formula proposed by Lenosky et al. in 1992 to describe the deformation energy of a single layer of curved graphite:[17]

$$E^g = \epsilon_0 \sum_{(ij)} \frac{1}{2}(r_{ij}-r_0)^2 + \epsilon_1 \sum_i \left( \sum_{(j)} \mathbf{u}_{ij} \right)^2 + \epsilon_2 \sum_{(ij)} (1-\mathbf{n}_i \cdot \mathbf{n}_j) + \epsilon_3 \sum_{(ij)} (\mathbf{n}_i \cdot \mathbf{u}_{ij})(\mathbf{n}_j \cdot \mathbf{u}_{ji}). \quad (1)$$

The first two terms are the contributions of bond length and bond angle changes to the energy. The last two terms are the contributions of the $\pi$-electron resonance. In the first term, $r_0 = 1.42$ Å is the initial bond length of planar graphite, and $r_{ij}$ is the bond length between atoms $i$ and $j$ after the deformations. In the remaining terms, $\mathbf{u}_{ij}$ is a unit vector pointing from atom $i$ to its neighbor $j$, and $\mathbf{n}_i$ is the unit vector normal to the plane determined by the three neighbors of atom $i$. The summation $\Sigma_{(j)}$ is taken over the three nearest-neighbor $j$ atoms to $i$ atom and $\Sigma_{(ij)}$ taken over all nearest-neighbor atoms. The parameters $(\epsilon_1, \epsilon_2, \epsilon_3) = (0.96, 1.29, 0.05)$ eV were determined by Lenosky et al.[17] through a local density approximation. In 1997, one of present authors and his co-workers[18] reduced Eq. (1) to a continuum limit form without taking the bond length change into account and obtained the curvature elastic energy of a SWNT,

$$E^{(s)} = \oint \left[ \frac{1}{2} k_c (2H)^2 + \bar{k}_1 K \right] dA, \qquad (2)$$

where the bending elastic constant

$$k_c = (18\epsilon_1 + 24\epsilon_2 + 9\epsilon_3) r_0^2 / (32\Omega) = 1.17 \text{ eV}, \qquad (3)$$

with $\Omega = 2.62$ Å$^2$ being the occupied area per atom, and

$$\bar{k}_1 / k_c = -(8\epsilon_2 + 3\epsilon_3)/(6\epsilon_1 + 8\epsilon_2 + 3\epsilon_3) = -0.645. \qquad (4)$$

In Eq. (2), $H$, $K$, and $dA$ are the mean curvature, Gaussian curvature, and area element of the SWNT surface, respectively. The obtained value of $k_c$ is in reasonable agreement with the value of 1.02 eV calculated by Tersoff[19] using a valence force model for straight tubes and is excellently close to the value of 1.2 eV extracted from the measured phonon spectrum of graphite.[20] The calculated ratio of $\bar{k}_1 / k_c$ is also close to the result of $\bar{k}_1 / k_c = -88/105.4 = -0.8$ measured by Blakeslee et al.[21]

In fact, the continuum form of the last three terms of Eq. (1) is still not complete to describe the mechanical properties of SWNT's. For a SWNT with in-plane deformations—e.g., the SWNT loaded with external forces—the contribution of the first term in Eq. (1) must be considered since the bond length change is remarkable in this case. In what follows, we generalize the derivation of Eq. (2) to involve the in-plane deformations; i.e., the continuum limit form of the first term in Eq. (1) will be incorporated into the energy expression of a deformed SWNT.

Let us first consider a SWNT with in-plane deformations,

$$\boldsymbol{\varepsilon}_i = \begin{pmatrix} \varepsilon_x & \varepsilon_{xy} \\ \varepsilon_{xy} & \varepsilon_y \end{pmatrix},$$

at the $i$-atom site, where $\varepsilon_x$, $\varepsilon_y$, and $\varepsilon_{xy}$ are the axial, circumferential, and shear strains, respectively. In other words, the bond vector $\mathbf{r}_{ij}$ from atom $i$ to its neighbor $j$ after the deformations and the initial bond vector $\mathbf{r}_{ij}^0$ before the deformations satisfy $\mathbf{r}_{ij} = (\mathbf{I} + \boldsymbol{\varepsilon}_i) \mathbf{r}_{ij}^0$,[16] where $\mathbf{I}$ is a $2 \times 2$ unit matrix. If regarding $\mathbf{r}_{ij}^0 = \mathbf{r}_j^0 - \mathbf{r}_i^0$ with $\mathbf{r}_i^0$ and $\mathbf{r}_j^0$ being the positions of atoms $i$ and $j$ on the surface of the SWNT before the deformations, then we can transform the initial bond vectors $\mathbf{r}_{ij}^0$ into the continuum limit by expanding them up to the order of $O(r_0^2 \kappa^2)$ as[22]

$$\mathbf{r}^0(M) = \mathbf{r}_{ij}^0 = [1 - r_0^2 \kappa^2(M)/6] r_0 \mathbf{t}(M) + [r_0 \kappa(M)/2 + r_0^2 \kappa_s(M)/6] r_0 \mathbf{N}(M) + [\kappa(M) \tau(M) r_0^2 / 6] r_0 \mathbf{b}(M), \qquad (5)$$

where $M = 1, 2, 3$ denote three $sp^2$-bonded curves from atom $i$ to one of its three neighbor atoms $j$ on the SWNT surface. The symbols $\mathbf{t}$, $\mathbf{N}$, and $\mathbf{b}$ represent the unit tangential, normal, and binormal vectors of the bond curves from $i$ atom to $j$ atom, and $\kappa$ and $\tau$ refer to the curvature and torsion of them. Here, $s$ is the arc length parameter along the bond curve and $\kappa_s = d\kappa/ds$. The vectors $\mathbf{t}(M)$ and $\mathbf{b}(M)$ can be expressed by $\mathbf{t}(M) = \cos \theta(M) \hat{\mathbf{x}} + \sin \theta(M) \hat{\mathbf{y}}$ and $\mathbf{b}(M) = -\sin \theta(M) \hat{\mathbf{x}} + \cos \theta(M) \hat{\mathbf{y}}$, where $\hat{\mathbf{x}}$ and $\hat{\mathbf{y}}$ are the unit axial and circumferential vectors at the $i$-atom site on the SWNT surface, and $\theta(M)$ is the rotating angle from $\hat{\mathbf{x}}$ to $\mathbf{t}$. Thus we have the expressions of $\mathbf{u}_{ij} = \mathbf{r}_{ij}/r_{ij}$ and $\mathbf{n}_i = \mathbf{N}(M)$ with $r_{ij} = |\mathbf{r}_{ij}|$ for the deformed SWNT. Considering $\Sigma_{M=1}^3 \sin^2 \theta(M) = \Sigma_{M=1}^3 \cos^2 \theta(M) = 3/2$ and $\Sigma_{M=1}^3 \sin^4 \theta(M) = \Sigma_{M=1}^3 \cos^4 \theta(M) = 9/8$ as well as the above expressions, and transforming Eq. (1) into continuum limit up to the second-order magnitudes of $\varepsilon_x$, $\varepsilon_y$, $\varepsilon_{xy}$, and $r_0 \kappa$, we find that the terms of Eq. (1) related to the curvatures naturally enter Eq. (2) and the remaining terms related to the in-plane deformations can be expressed as[23]

$$E_d = \oint \left[ \frac{1}{2} k_d (2E)^2 + \bar{k}_2 F \right] dA, \qquad (6)$$

where $E = (\varepsilon_x + \varepsilon_y)/2$ and $F = \varepsilon_x \varepsilon_y - \varepsilon_{xy}^2$ are, respectively, named "mean" and "Gaussian" strains, and

$$k_d = 9(\epsilon_0 r_0^2 + \epsilon_1)/(16\Omega), \qquad (7)$$

$$\bar{k}_2 = -3(\epsilon_0 r_0^2 + 3\epsilon_1)/(8\Omega). \qquad (8)$$

The value of $\epsilon_0$ was not given by Lenosky et al.,[17] but Zhou et al.[6] gave $\epsilon_0 = 57$ eV/Å$^2$ from the force-constant method. Thus from Eqs. (7) and (8), we get $k_d = 24.88$ eV/Å$^2$ and $\bar{k}_2/k_d = -0.678$ which is excellently close to the value of $\bar{k}_1/k_c$ shown in Eq. (4). Therefore, we can regard both $\bar{k}_2/k_d$ and $\bar{k}_1/k_c$ as equal to their average value,

$$\bar{k}_1/k_c = \bar{k}_2/k_d = -0.66. \qquad (9)$$

This is the key relation that allows us to describe the deformations of a SWNT with classic elastic theory. Thus, the deformation energy of a SWNT, the sum of Eqs. (2) and (6),

$$E_d^{(s)} = \oint \left[ \frac{1}{2} k_c (2H)^2 + \bar{k}_1 K \right] dA + \oint \left[ \frac{1}{2} k_d (2E)^2 + \bar{k}_2 F \right] dA, \qquad (10)$$

can be expressed as a form of the classic shell theory:[3–5,24]

TABLE I. Some typical values of Young's modulus $Y$ (in unit TPa), Poisson ratio $\nu$, and effective wall thickness $h$ (in units of Å) of single-walled carbon nanotubes given by different authors. The last column lists the values of Young's modulus per atom $Y_{atom}$ (in unit eV/atom).

| Authors | $Y$ | $\nu$ | $h$ | $Y_{atom}$ |
| --- | --- | --- | --- | --- |
| Yakobson et al.[a] | 5.5 | 0.19 | 0.66 | 59.4 |
| Yakobson et al.[b] | 3.859 | 0.149 | 0.894 | 57.3 |
| Zhou et al.[c] | 5.1 | 0.24 | 0.74 | 61.8 |
| Lu[d] | 1.0 | 0.28 | 3.4 | 55.7 |
| Present authors | 4.7 | 0.34 | 0.75 | 57.7 |

[a]References 3 and 4.
[b]Reference 5.
[c]Reference 6.
[d]Reference 8.

$$E_c = \frac{1}{2} \oint D[(2H)^2 - 2(1-\nu)K]dA$$
$$+ \frac{1}{2} \oint \frac{C}{1-\nu^2}[(2E)^2 - 2(1-\nu)F]dA, \quad (11)$$

where $D=(1/12)Yh^3/(1-\nu^2)$ and $C=Yh$ are the bending rigidity and in-plane stiffness of the shell. $\nu$ is the Poisson ratio and $h$ is the thickness of the shell. Comparing Eqs. (10) and (11), we have

$$(1/12)Yh^3/(1-\nu^2) = k_c,$$
$$Yh/(1-\nu^2) = k_d, \quad (12)$$
$$1-\nu = -\bar{k}_1/k_c = -\bar{k}_2/k_d.$$

From above equations we obtain the Poisson ratio, effective wall thickness, and Young's modulus of SWNT's as $\nu = 0.34$, $h = 0.75$ Å, and, $Y = 4.70$ TPa, respectively. Our numerical results are close to those given by Yakobson and co-workers[3–5] and Zhou et al.[6] Some typical data given by different authors are listed in Table I, from which we find that different authors give the approximately consistent values of Young's modulus per atom $Y_{atom}$ defined by $Y\Omega h$, though they each give inconsistent values of $Y$ and $h$. In the recent literature, some authors proposed to adopt $Y_s = Yh$ as the Young's modulus of a SWNT.[7] This definition is appropriate only in dealing with the pure stretch and compression deformations of SWNT's where the strain energy depends on $Yh$, but is not consistent with the bending deformation of a SWNT where the strain energy depends on $Yh$ and $Yh^3$.[6] Therefore, it is necessary to determine $Y$ and $h$ separately.

We now turn to consider a deformed MWNT. For simplicity, we merely consider a straight MWNT loaded with uniform axial stresses at its two ends—i.e., the case of $K=0$ and $\varepsilon_{xy}=0$. As shown in Fig. 1, a MWNT can be thought of as a group of the above-mentioned SWNT's with a common center axis. $\rho_i$ and $\rho_o$ are the inmost and outmost radii of the MWNT. $h = 0.75$ Å is the effective wall thickness of each SWNT and $d = 3.4$ Å is the distance between layers of the MWNT. The number of layers, $N$, is given by $N=(\rho_o - \rho_i)/d + 1$ and the $l$th layer radius is $\rho_l = \rho_i + (l-1)d$ ($l = 1, 2, \ldots, N$). Then the energy of a free MWNT is obtained as[18]

$$E^{(m)} = \sum_{l=1}^{N} \pi k_c L/\rho_l - \sum_{l=1}^{N-1} g\pi L(\rho_{l+1}^2 - \rho_l^2), \quad (13)$$

with $g \sim -\Delta E_{coh}/d$ and $\Delta E_{coh} = -2.04$ eV/nm² being the interlayer cohesive energy of 1 nm² area of planar graphite obtained theoretically by Girifalco and Lad.[25] The first term in Eq. (13) expresses the summation of curvature energies on all layers given in Eq. (2), and the second term represents the total interlayer cohesive energy which actually arises from the relatively weaker van der Waals interactions.[8,18] On this account, we can reasonably believe that the axial strain $\varepsilon_x$ and circumferential strain $\varepsilon_y$ still satisfy $\varepsilon_y = -\nu\varepsilon_x$ for every layer of SWNT's in the MWNT. All layers have the same circumferential strain, since they have the same axial strain after the uniform stresses loading on the two ends of the MWNT. From complicated calculations,[23] we find that the energy variation caused by the curvature and volume changes can be neglected as compared with that caused by in-plane deformations. Thus the energy variation up to the second-order magnitudes of $\varepsilon_x$ and $\varepsilon_y$ can be derived as

$$\Delta E^{(m)} = \frac{k_d}{2}(1-\nu^2)\varepsilon_x^2 \sum_{l=1}^{N} 2\pi\rho_l L. \quad (14)$$

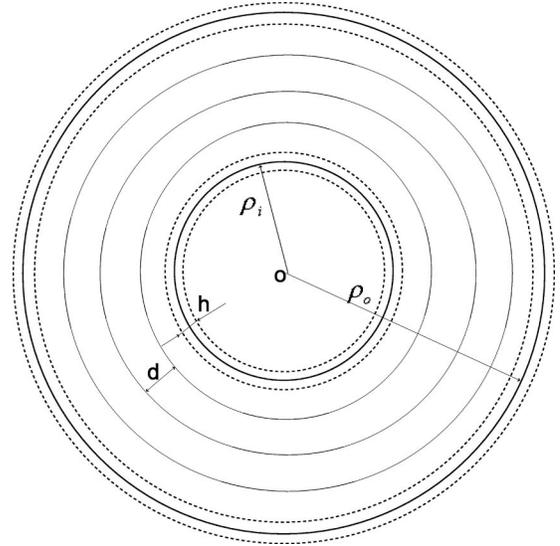

FIG. 1. The construction of a MWNT cross section perpendicular to the tube axis. A MWNT can be thought of as a group of SWNT's with a common center axis. $\rho_i$ and $\rho_o$ are the inmost and outmost radii of the MWNT. $h$ is the effective wall thickness of a SWNT and $d$ is the distance between layers of the MWNT.

The effective Young's modulus of the MWNT $Y_m$ is defined as the second-order partial differential of the energy variation to the axial strain in the unit volume,

TABLE II. The relation between Young's moduli $Y_m$ (in unit TPa) and the layer number $N$ of multiwalled carbon nanotubes. $N=1$ corresponds to the case of single-walled carbon nanotubes.

| $N$ | 1 | 2 | 3 | 4 | 5 | 8 | 10 | 20 | 100 |
|---|---|---|---|---|---|---|---|---|---|
| $Y_m$ | 4.70 | 1.70 | 1.41 | 1.29 | 1.23 | 1.15 | 1.13 | 1.08 | 1.05 |

$$Y_m = \frac{1}{V}\frac{\partial^2 \Delta E^{(m)}}{\partial \varepsilon_x^2}, \quad (15)$$

where $V = \pi L[(\rho_o + h/2)^2 - (\rho_i - h/2)^2]$ is the volume of the MWNT (see also Fig. 1). Considering $\Sigma_{l=1}^{N} 2\pi\rho_l L = (\rho_i + \rho_o)N\pi L$ and substituting Eqs. (12) and (14) into Eq. (15), we obtain

$$Y_m = \frac{N}{N-1+h/d}\frac{h}{d}Y, \quad (16)$$

where $Y$ and $h$ are Young's modulus and effective wall thickness of the SWNT's. Obviously, $Y_m = Y = 4.70$ TPa if $N=1$, which corresponds to the result of SWNT's, and $Y_m = Yh/d = 1.04$ TPa if $N \gg 1$, which is just the Young's modulus of bulk graphite. The values of the effective Young's moduli $Y_m$ and the corresponding layer number $N$ are listed in Table II which suggests that the Young's moduli of MWNT's depend on the number of tube layers and the varying range of moduli is very large (from 4.70 to 1.04 TPa for $N=1-\infty$). In fact, the value of $Y_m$ does not reflect the physics change in the true lattice rigidity but just a choice of the cross section. This aspect has been discussed carefully in a recent review by Yakobson and Avouris.[4] But the Young's moduli of MWNT's tested by experiments are usually the effective moduli. Thus $Y_m$ is useful to the experiments and the large divergence[9,10] in measuring the Young's moduli of MWNT's may be understandable.

In summary, by reducing Eq. (1) into the continuum limit form, we show that the classic shell theory can serve to describe the deformations of SWNT's and obtain the Poisson ratio, the effective thickness, and the Young's modulus of SWNT's. Based on the results, we theoretically reveal the effective Young's moduli of MWNT's to be dependent on their layer number and with a large varying range. There is an open question on the microscopic mechanism of elastic moduli, which still needs further discussions. It is necessary to point out that our conclusions are valid only for tubes whose radii are not very small because of expanding Eq. (1) up to $O(r_0^2 \kappa^2)$; therefore, to deal with the Young's moduli of extremely small CNT's will be our future work.

The authors acknowledge the useful discussions with Professor H. W. Peng, Dr. X. Zhou, M. Li, and J. J. Zhou, and the warm help of Professor B. I. Yakobson.